\documentclass[aps,prl,twocolumn,superscriptaddress]{revtex4}

\usepackage{amsmath}
\usepackage{amssymb}
\usepackage{graphicx}
\usepackage{array}
\usepackage{siunitx}
\usepackage[usenames, dvipsnames]{color}

\begin{document}
\title{Energy-chirp compensation in laser wakefield accelerators}

\author{A. D\"opp}
\affiliation{LOA, ENSTA ParisTech - CNRS - \'Ecole Polytechnique - Universit\'e Paris-Saclay, 828 Boulevard des Mar\'echaux, 91762 Palaiseau Cedex, France }
\affiliation{Ludwig-Maximilians-Universit\"at M\"unchen, Am Coulombwall 1, 85748 Garching, Germany}
\author{C. Thaury}
\affiliation{LOA, ENSTA ParisTech - CNRS - \'Ecole Polytechnique - Universit\'e Paris-Saclay, 828 Boulevard des Mar\'echaux, 91762 Palaiseau Cedex, France }
\author{E. Guillaume}
\affiliation{LOA, ENSTA ParisTech - CNRS - \'Ecole Polytechnique - Universit\'e Paris-Saclay, 828 Boulevard des Mar\'echaux, 91762 Palaiseau Cedex, France }
\author{F. Massimo}
\affiliation{LOA, ENSTA ParisTech - CNRS - \'Ecole Polytechnique - Universit\'e Paris-Saclay, 828 Boulevard des Mar\'echaux, 91762 Palaiseau Cedex, France }
\author{A. Lifschitz}
\affiliation{LOA, ENSTA ParisTech - CNRS - \'Ecole Polytechnique - Universit\'e Paris-Saclay, 828 Boulevard des Mar\'echaux, 91762 Palaiseau Cedex, France }
\author{I. Andriyash}
\affiliation{LOA, ENSTA ParisTech - CNRS - \'Ecole Polytechnique - Universit\'e Paris-Saclay, 828 Boulevard des Mar\'echaux, 91762 Palaiseau Cedex, France }
\affiliation{Department of Physics and Complex Systems, Weizmann Institute of Science, Rehovot, 76100, Israel}
\author{J.-P. Goddet}
\affiliation{LOA, ENSTA ParisTech - CNRS - \'Ecole Polytechnique - Universit\'e Paris-Saclay, 828 Boulevard des Mar\'echaux, 91762 Palaiseau Cedex, France }
\author{A. Tazfi}
\affiliation{LOA, ENSTA ParisTech - CNRS - \'Ecole Polytechnique - Universit\'e Paris-Saclay, 828 Boulevard des Mar\'echaux, 91762 Palaiseau Cedex, France }
\author{K. Ta Phuoc}
\affiliation{LOA, ENSTA ParisTech - CNRS - \'Ecole Polytechnique - Universit\'e Paris-Saclay, 828 Boulevard des Mar\'echaux, 91762 Palaiseau Cedex, France }
\author{V. Malka}
\affiliation{LOA, ENSTA ParisTech - CNRS - \'Ecole Polytechnique - Universit\'e Paris-Saclay, 828 Boulevard des Mar\'echaux, 91762 Palaiseau Cedex, France }
\affiliation{Department of Physics and Complex Systems, Weizmann Institute of Science, Rehovot, 76100, Israel}

\begin{abstract}

The energy spread in laser-wakefield accelerators is primarily limited by the energy-chirp introduced during the injection and acceleration processes. Here we propose and demonstrate the use of longitudinal density tailoring to adapt the accelerating fields and reduce the chirp at the end of the accelerator. Experimental data supported by 3D PIC simulations show that broadband electron beams can be converted to quasi-monoenergetic beams of $\leq10\%$ energy spread while maintaining a high charge of more than 120 pC. In the linear and quasi-linear regimes of wakefield acceleration, the method could provide even lower, sub-percent level, energy spread.
 
\end{abstract}
\maketitle


\section{Introduction}

Laser wakefield acceleration is an aspiring technology to produce femtosecond bunches of highly relativistic electrons in a compact way \cite{Malka:2008un,Esarey:2009ks,Hooker:2013jk}. While the high field gradients permit acceleration of electrons to hundreds of megaelectronvolt on a millimeter scale \cite{Malka:2002eu}, they also cause a large energy spread because of the different accelerating fields experienced by electrons within a bunch. In many cases the final energy spread is of the order of the momentum difference between those particles that are trapped and accelerated at earlier and later times.

Hence, one obvious way to reduce the energy spread is to reduce the injection length. Indeed, if the trapping conditions are only met during a short moment, quasi-monoenergetic electron beams are observed \cite{Mangles:2004vr,Geddes:2004vs,Faure:2004tj}. To enter this regime, the accelerator is operated at a relatively low plasma density so that injection relies on the laser propagation, when self-focusing and pulse-compression trigger an expansion of the plasma cavity \cite{Kalmykov:2009id}. Once a certain charge has been trapped inside the wake, beam-loading prohibits further injection \cite{Benedetti:2013fy}. 

However, if injection relies on the non-linear laser propagation, stability and tunability are difficult to achieve \cite{Dopp:2017dza}. In response to this challenge, a number of controlled injection schemes have been developed to provide favorable trapping conditions at a defined time and position \cite{Malka:2012bi}. The most prominent techniques are colliding-pulse injection \cite{Faure:2006vy} and shock-front injection \cite{Schmid:2010ih}, which allow for accurate tuning of the electron beam energy while maintaining a low energy spread. But electron beams from such localized injection schemes typically contain an order of magnitude less charge than broadband beams from self-injection, ionization-induced injection \cite{Clayton:2010kh,Guillaume:2015gy} or density downramp injection \cite{Bulanov1998,Geddes:2008tj}. 

A less explored alternative for reducing the electron injection lenght is energy chirp compensation. In a laser wakefield accelerator, a plasma wave is formed behind the laser pulse which propagates at the group velocity $v_g$. In contrast, highly relativistic electrons with a Lorentz factor $\gamma\gg1$ propagate at nearly the speed of light in vacuum ($v_e\simeq c_0$) and will gain on the laser beam and its wake during the acceleration process. When new electrons get subsequently injected at the back of the wake, this results in a clear relation between position and energy. Initially, this relation can be described as a linear chirp $\alpha$ and the momentum spread $\sigma_{p_z}\simeq \Delta \gamma$ is determined by the bunch length $\sigma_z$ according to $\sigma_{p_z} =  \sigma_z\times |\alpha|$. 

\begin{figure}[b]
\centering
\includegraphics[width=.9\linewidth]{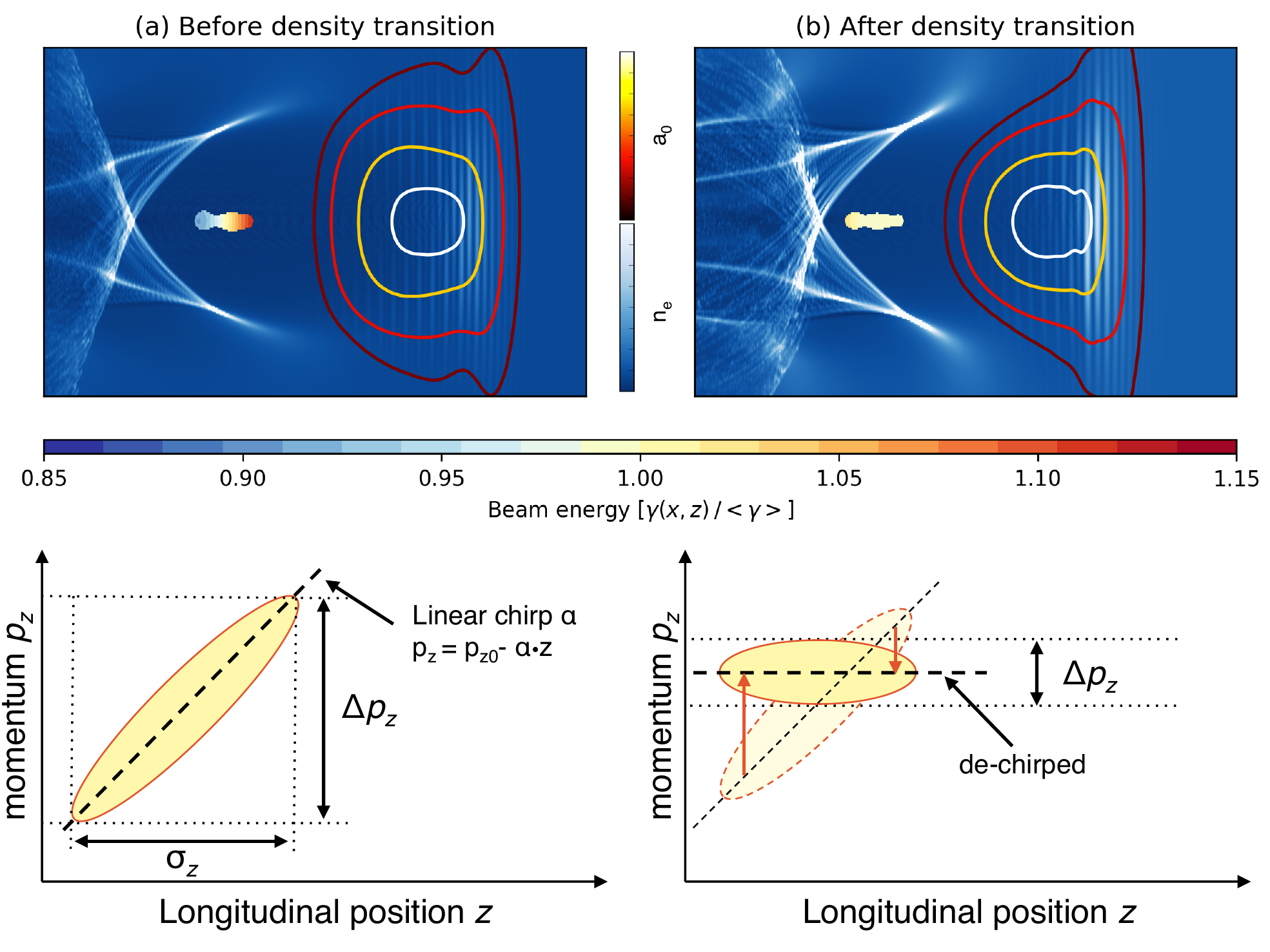}
\caption{Illustration of chirp reduction as a result of a density transition. Upper part: plasma density (blue colormap), laser intensity (isolevels) and beam energy before (a) and after (b) a density transition as calculated with PIC simulations. Lower part: Sketch of the  $(z,p_z)$ phase space for both cases. The beam is initially chirped (dashed line) and therefore electrons of different energy are located at different phases of the wakefield. Using the density transition, the phase space ellipse (yellow) can be rotated, thus reducing both chirp and beam energy spread.}
\label{Fig1}
\end{figure}

\begin{figure*}[t]
\centering
\includegraphics[width=0.95\linewidth]{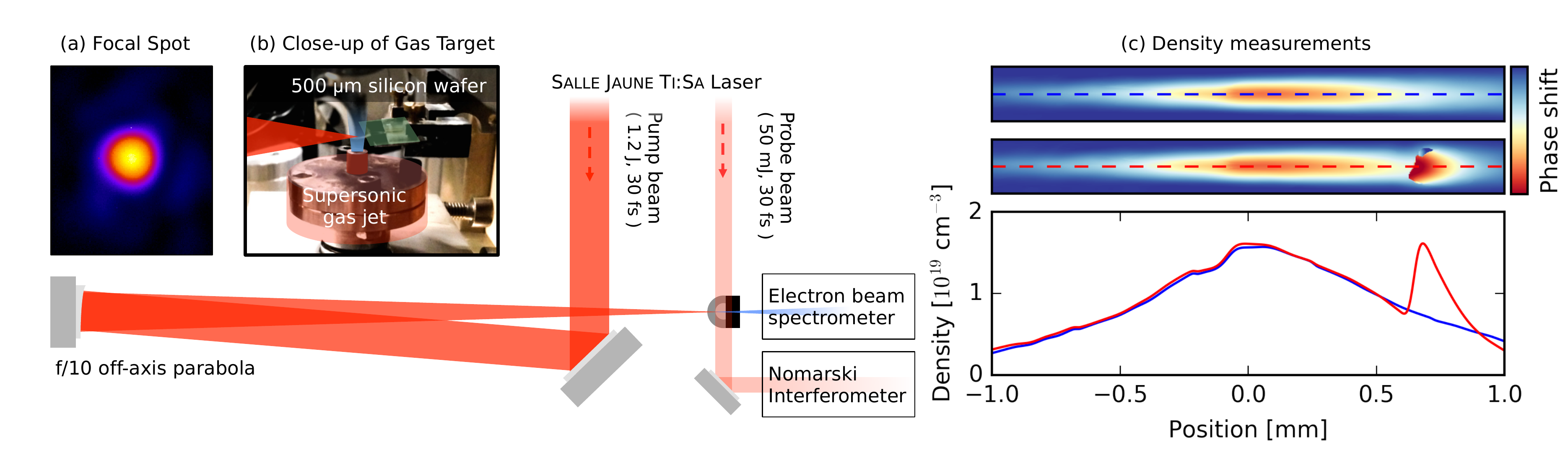}
\caption{Sketch of the experimental setup and results from density measurements. The density transition is induced using a silicon wafer that is moved into the flow of the supersonic gas jet, see inset (b). Upper part in (c): Phase maps retrieved for a case without (top) and with shock (bottom). Lower part in (c): Reconstructed density profiles using Abel inversion along the rotation axes indicated in the upper plot.}
\label{Fig2}
\end{figure*}

However, the longitudinal wakefield is not constant, but changes from accelerating in the rear part to decelerating in the front. Once electrons are injected and advance with respect to the wake ($v_e>v_g$), their phase no longer matches the ideal accelerating phase. This process is called dephasing and it is mostly known as a major limiting factor for the maximum energy gain in laser wakefield accelerators. A side effect of dephasing is that it can reduce the electron energy spread: During the dephasing process, electrons at the bunch head start to decelerate, while the other electrons still gain energy. If the accelerator length is tuned closely to the dephasing length $L_d$, the linear chirp is naturally compensated. Experimentally, such alignment can be achieved by changing the background plasma density, but this is undesirable because the plasma density will also affect laser propagation, plasma wake formation and electron injection. This drawback is avoided with gas cells \cite{Corde:2013gj,Heigoldt:2015cd}, whose length can be adjusted at constant plasma density in order to match the dephasing length. Yet this will only compensate the chirp for a fixed beam energy and accelerator length. If the acceleration is stopped before, e.g. if the dephasing length is longer than the pump depletion length or the effective Rayleigh length, the electron beam spread remains increased due to the non-zero chirp. 

Recently, it has been shown how longitudinal density tailoring can be used to influence the laser wakefields \cite{Lehe:2014efa,Dopp:2016gm}. In particular, experiments using such profiles demonstrated reduced beam divergence \cite{Thaury:2015cg} and increased beam energy \cite{Guillaume:2015dia}. Here, we discuss how a similar approach can be used to manipulate the beam chirp, and thus to control the energy spread, at the end of the accelerator. 

The basic principle of the density-induced chirp compensation is illustrated in Fig.1. Most injection schemes such as self-injection, downramp injection or ionization-induced injection result in electron beams with negative chirp. However, a transition to higher plasma density can be used to increase the acceleration of the rear part of the bunch, while the front is less accelerated and later decelerated. The beam therefore rotates in the $(z,p_z)$ phase space, as shown in the lower part of Fig.1, and its chirp and energy spread are reduced \footnote{These qualitative simulations have been performed using the Particle-In-Cell Code \textsc{Osiris}, using a 2-D cartesian grid with a resolution of $\Delta x = 0.32k_0^{-1}$ and $\Delta y = 1.5 k_0^{-1}$ and the time step is $c\Delta t = 0.9\Delta x$. The plasma is initialized with 2 particles per cell, which are interpolated using quadratic interpolation. The laser pulse with peak potential $a_0=4.0$ is focused at matched spot size $k_p w_0=2 \sqrt a_0$ into a plasma with density $n_0=5\times 10^{18}$ cm$^{-3}$. The electron beam profile is modeled with as a third-order super-gaussian in $x$ ($\sigma_x=0.5k_p^{-1}$) and second-order super-gaussian in $y$ ($\sigma_y=0.25k_p^{-1}$), with 10 particles per cell. The peak density of the beam is $n_b=2n_e$, resulting in weak beam-loading at the location of the electrons. In order to inject electron into the wake, the electrons are initialized with an initial longitudinal momentum of 20 MeV plus 45 MeV/\SI{}{\micro\metre}, corresponding to a linear chirp of about -15 MeV/\SI{}{\femto\second}. This degree of chirp is comparable to what is observed in 3-D PIC simulations.}.  We note that a similar method was theoretically studied by Hu et al.\cite{Hu:2016kw}

\section{Experimental results}

We have demonstrated the method experimentally with the 40 TW \textsc{Salle Jaune} Ti:Sa laser system at Laboratoire d'Optique Appliqu\'ee, which delivers pulses of 1.2 J energy on target with 30 fs duration and 800 nm central wavelength. The laser is focused at the entrance of a supersonic Helium gas jet using a f/10 off-axis parabola. The focal spot size is 15 \SI{}{\micro\metre}, with a peak intensity on target of $I = 1.0 \times 10^{19}$ W/cm$^{2}$, which corresponds to a normalized vector potential $a_0 = 2.2$. 

As in Ref.\cite{Guillaume:2015dia}, in-situ plasma density measurements are obtained from the phase shift of a low-energy probe beam, which propagates perpendicular to the drive beam through the target. The phase shift is measured using a Nomarski-type interferometer \cite{Benattar:1979uha} and the interference fringes are detected on a CCD camera at a resolution of 3.8 \SI{}{\micro\metre} per pixel. The fringe distance in the reference without plasma is $\sim14$ pixel, which limits the spatial resolution to approximately 50 \SI{}{\micro\metre}. 
For each target configuration, we average the interferograms over 10 shots. Phase maps are recovered from these interferograms using a Continuous Wavelet Transform algorithm. Under the assumption that the plasma channel is symmetric around the laser propagation axis, the phase shift per unit volume can be calculated by Abel inversion and the local electron density $n_e$ is then deduced applying the dispersion law for cold plasmas ($\epsilon = 1 - n_e/n_c$). The results from this analysis is shown in Fig.\ref{Fig1}. It should be noted that the phase maps are not entirely symmetrical close to the shock front. This leads to an overestimate of the gradient length and an underestimate of the peak transition density. As shown in Fig.\ref{Fig1}, the density profile consists of a linearly rising and then falling slope of about 1 mm length each, peaking at $1.6\times 10^{19}$ cm$^{-3}$.

\begin{figure}[t]
\centering
\includegraphics[width=.95\linewidth]{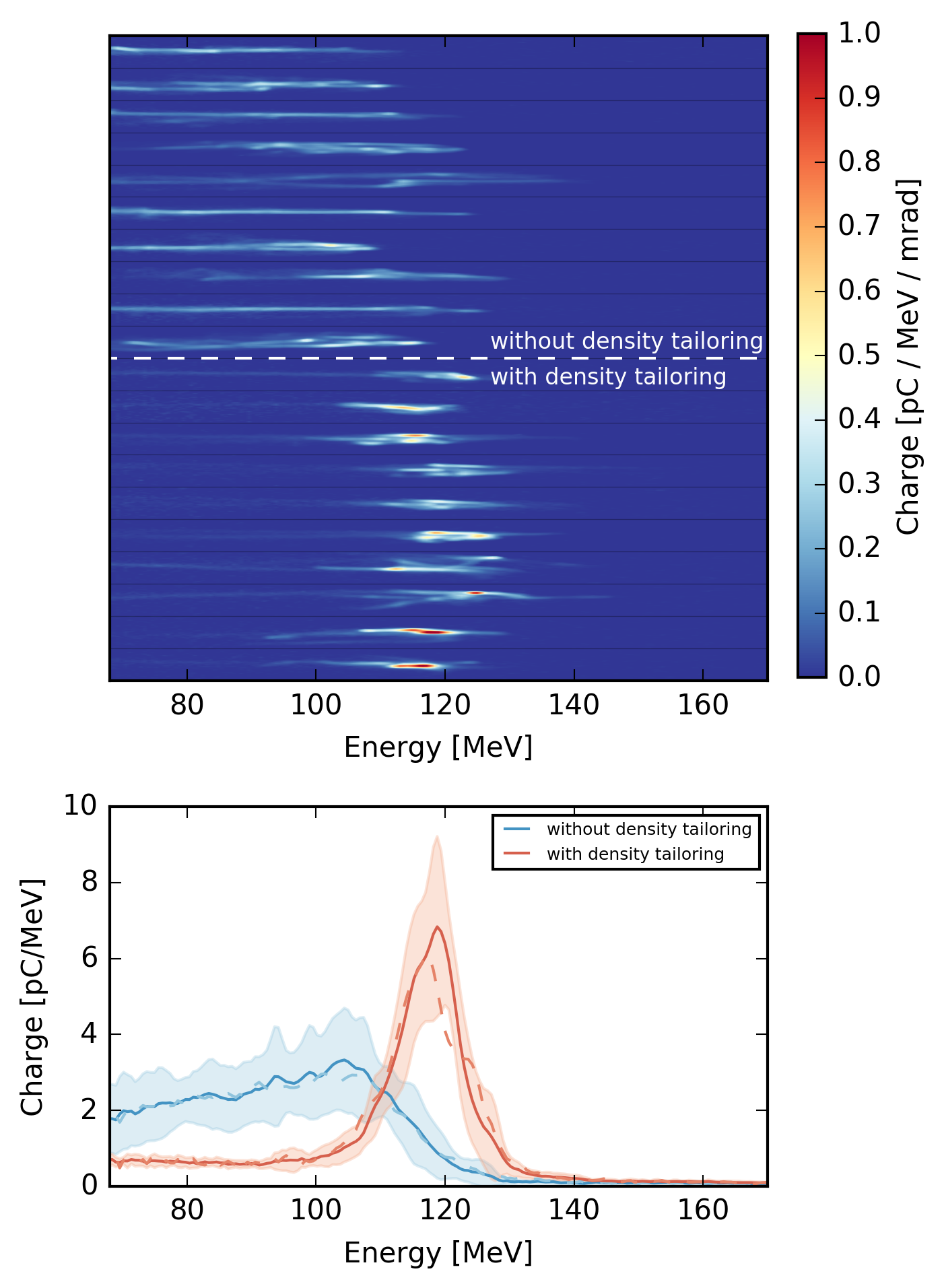}
\caption{Experimental data. (a) Density profile measurements. (b) Angularly resolved single-shot electron spectra for an unperturbed jet (above) and with density tailoring using a shock front (below). (c) Integrated electron spectra for both cases (dash lines for average spectra, solid lines for average spectra with corrected peak energies).}
\label{Fig3}
\end{figure}

While the laser propagates through the first part of the jet, the increasing plasma density causes the phase velocity of the wake to augment as well \cite{Faure:2010jk}. This should prevent injection at this stage of the interaction. In contrast, from the middle of the jet on, electrons are expected to be injected via density gradient injection \cite{Geddes:2008tj}. In accordance with this, the measured electron beams are spectrally broadband (see upper part of Fig.\ref{Fig2}). Here, electron beams are characterized with a magnet spectrometer, giving information about electron beam charge, divergence and and their energy spectrum from 70 MeV onwards. While the density downramp allows electrons to get trapped easily, it also reduces the effective acceleration length and field. As a result, the measured cut-off energy of $122\pm9$ MeV is lower than what would be expected for a flat density profile at the same peak density ($\sim 200$ MeV using the scalings from Lu et al. \cite{Lu:2007eb}). The beam charge is $146\pm22$ pC, with a divergence of $8\pm3$ mrad. Due to the broadband nature of the injection, the spectral charge density typically remains below 3 pC/MeV.

To create a density transition, a silicon wafer is used, which is mounted on a motorized stage at the rear part of the jet. The obstacle in the supersonic gas flow leads to the formation of a shock front that travels downstream \cite{Schmid:2010ih}. Placed at the leaving side of the jet, this results in a sharp upward density transition along the laser axis of propagation. The longitudinal position of this transition can be adjusted by moving the blade. While the density at the shock is similar to the density at the center of the jet, the plasma density rapidly decreases behind the shock, terminating the acceleration process.

As the density transition is introduced the electron energy distribution beam drastically changes (see Fig.2). With the transition located at $z_{transition}=0.7$ mm behind the center of the jet, the broad energy spectrum is converted into a distribution that peaks at $117\pm12$ MeV, with an energy spread of less than 10 percent, cf. Fig.2 bottom. The beam charge is similar, but slightly lower than in to the non-perturbed case ($123\pm18$ pC) and the spectral charge density at the peak increases to over 6 pC/MeV. The beam divergence remains unaffected ($8\pm3$ mrad). As expected, the final beam spectrum is sensitive to the position of the density transition. When the transition occurs too early, it disturbs the electron injection process and the electron beam is essentially lost. The further the silicon wafer is moved outside of the jet, the less pronounced the narrowing of the spectrum becomes, until the electron distribution resembles the case without density tailoring.

\begin{figure*}[pt]
\centering
\includegraphics[width=1.\linewidth]{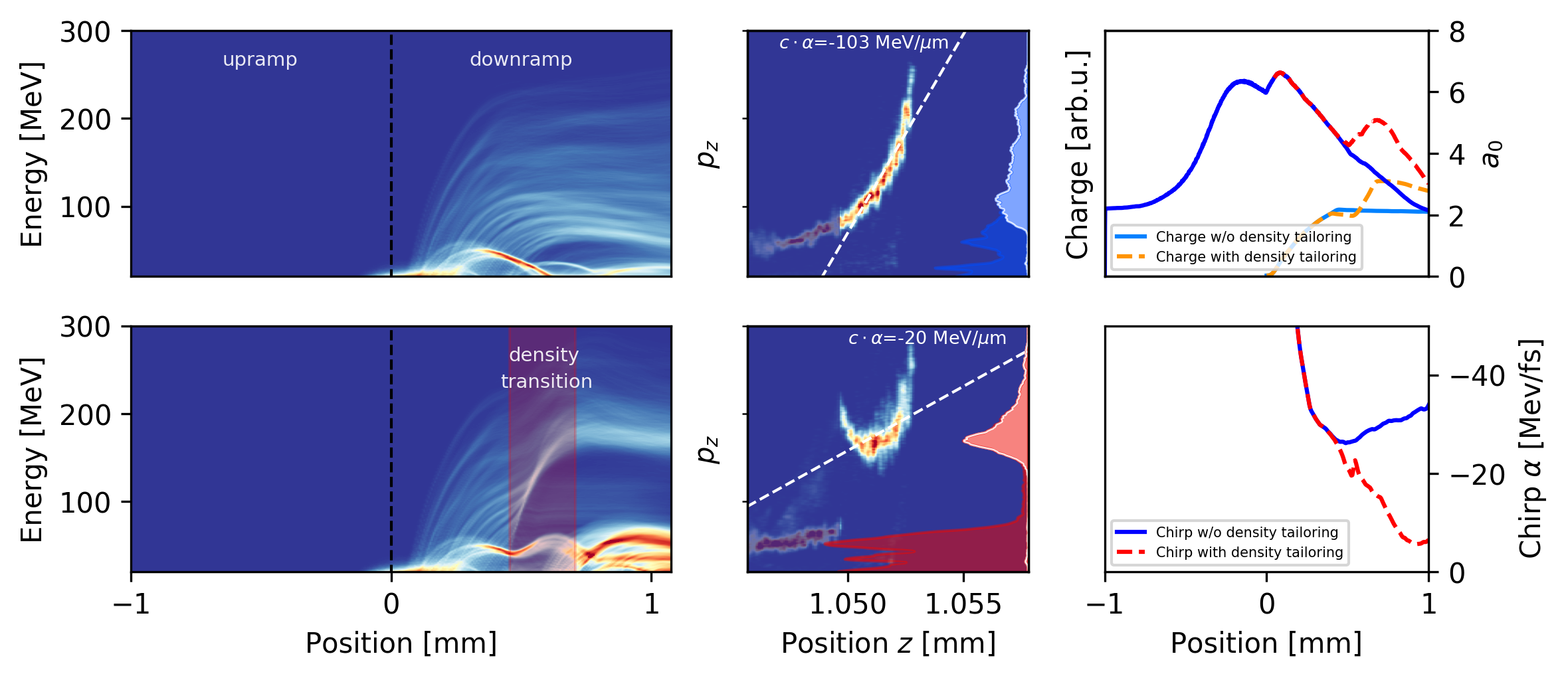}
\caption{Results from 3-D PIC simulations. Left: Evolution of the electron beam spectrum in the laser-wakefield accelerator without density tailoring (top) and with density tailoring (bottom). The corresponding $(z,p_z)$ phase spaces at the end of the accelerator are shown in the panel to the right. Light colors show the integrated spectrum within the rephasing region, shaded colors show the spectrum of the entire beam. Injection and laser dynamics are plotted in the top right, the evolution of the beam energy chirp is shown on the bottom right.}

\label{Fig3}
\end{figure*}
\section{Particle-In-Cell simulations}

To gain more insight in the physics that lead to this result, the experiment is modeled using the quasi-3D PIC code \textsc{Calder-Circ} \cite{Lifschitz20091803}. 	According to the experiment, a gaussian laser pulse is initialized at $z=-1$mm with $w_0 = 15$ \SI{}{\micro\metre}, $\tau =30$ fs and $a_0 = 2.2$, while the plasma density profile is defined from the experimentally measured profiles, with a peak density of $1.6\times 10^{19}$ cm$^{-3}$ at $z=0$ mm. The resolution is $\Delta z=0.25 k_0^{-1}$, $\Delta r=1.0 k_0^{-1}$ and $c\Delta t = 0.94 \Delta z $, with two Fourier modes ($m=0-1$) in the poloidal direction and 50 particles per cell. The results are summarized in Fig.3.

First, no electrons are accelerated during propagation along the density upramp. At the same time, the laser is self-focusing and self-compressing, reaching a peak vector potential $a_0=6.6$ in vicinity of the density peak at $z=0$ mm. Once the laser enters the downramp, the wakefield starts to expand and electrons are trapped and accelerated inside the bubble. However, injection in this regime is continuous and leads to a large energy spread. As observed in the experiment, the density-tailored case significantly changes the resulting electron spectra, exhibiting a clear peak at 168 MeV. The energy spread within the same beam region reduces from of 139 MeV at full-width at half-maximum to 39 MeV.

The simulations show that this behavior is primarily caused by a reduction in the energy chirp of the beam. As shown in Fig.3, the electron beam from downramp injection is strongly chirped. In the rephasing region, the linear chirp is more than 30 MeV/fs. We find that the linear chirp is significantly reduced with the density transition, to less than 10 MeV/fs. As conceived, this is the result of enforced acceleration fields at the back of the laser wakefield. While this leads to nearly chirp-free regions in the center of the beam, the field structure is not ideal and the bunch head and tail have a higher energy, leading to non-linear chirp components. 

Another observation of the experiments is that the bunch charge above the detection threshold of 70 MeV is comparable for both cases. Indeed, when considering the charge above this threshold, the density-tailored case yields 88 percent of the unperturbed charge, which is very similar to the experiment (84 \%). This is significantly higher than in first demonstrations of electron rephasing with self-injected beams \cite{Guillaume:2015dia}. While a part of the low-energy tail is still lost during the density upramp, the charge below 70 MeV is even higher with density tailoring, which is due to injection in the steep density downramp at the rear side of the shock front.

\begin{figure}[t]
\centering
\includegraphics[width=1.0\linewidth]{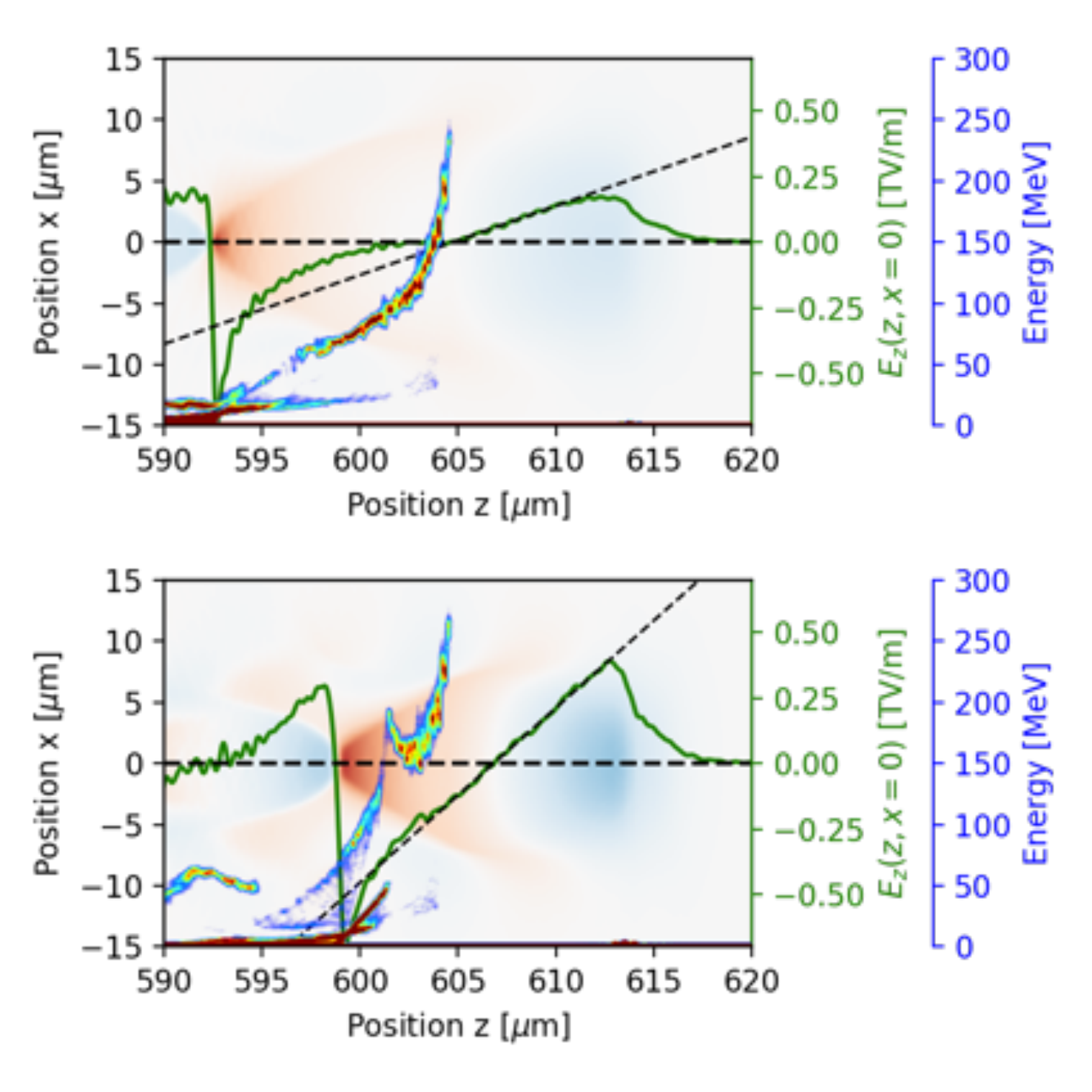}
\caption{Beam-loading in the density-tailored accelerator. Longitudinal wakefields $E_z$ and electron $(z-p_z)$ phase space in the case without density tailoring (top) and with density tailoring (bottom). The dashed line shows a linear fit to the wakefield, which approximately corresponds to the situation in an unloaded cavity.}
\label{FIG5}
\end{figure}

The simulation results support the conclusion that the observed energy spread reduction is due to a significant reduction in energy-chirp. Importantly, the effect is induced by changes of the plasma density and other effects such as beam-loading and the laser evolution play only a minor role in the process. For instance, our simulations show weak beam-loading at the location of the electron beam in the unperturbed case (see Fig.\ref{FIG5}). Beam-loading scales with the ratio of the beam density $n_b$ to the background plasma density $n_e$ and accordingly, the effect is suppressed at the higher plasma densities of the density-tailored case. Note that due to the potential loss of electrons injected shortly before the density transition, it is preferable to compensate the beam chirp once the entire beam has advanced inside the ion cavity (as in Fig.1). 

\begin{figure}[b]
\centering
\includegraphics[width=1.\linewidth]{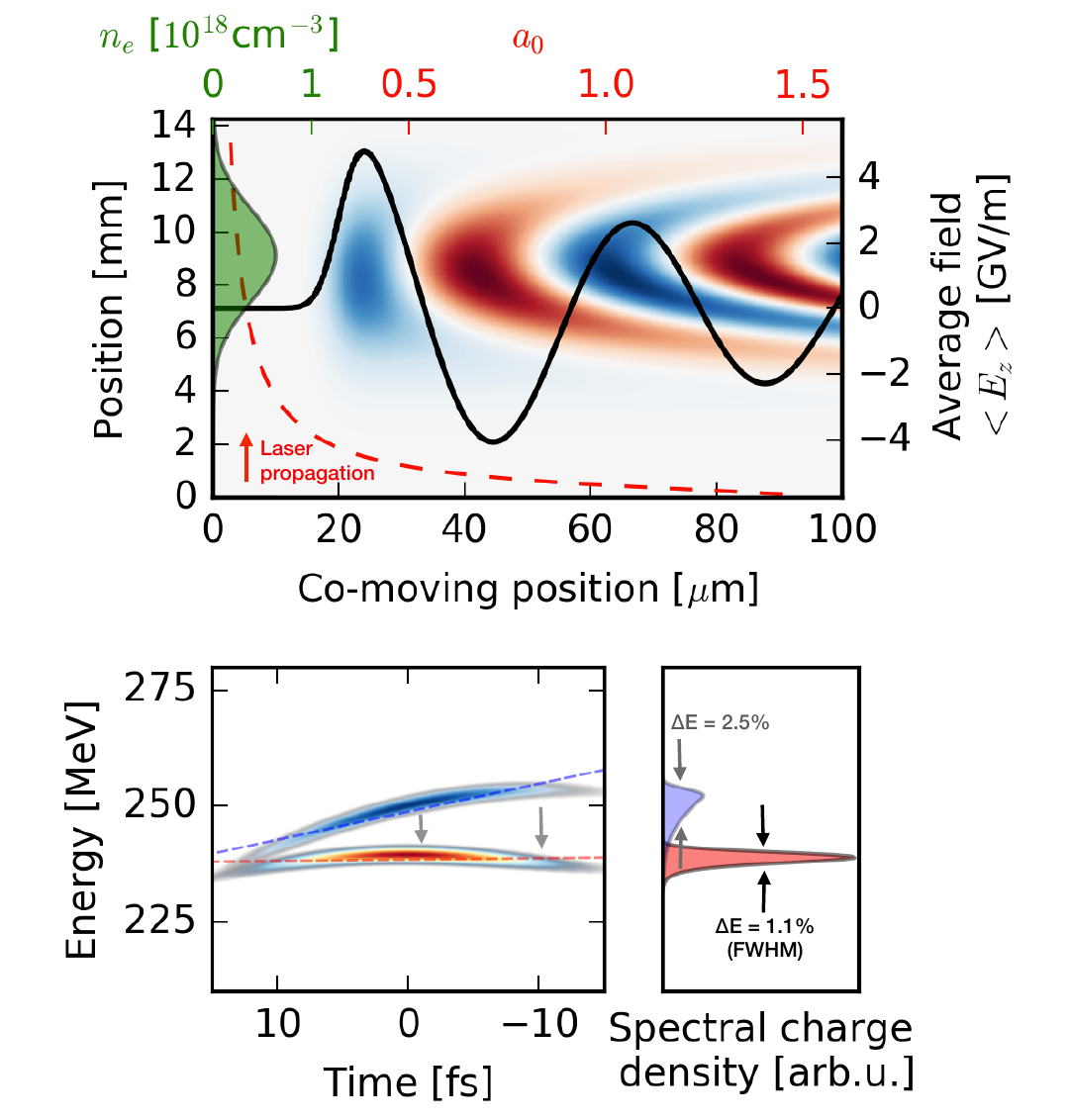}
\caption{Simulated chirp compensation in the linear wakefield regime. Upper plot: Wakefield (colormap) and average electric field (black line) in the wakefield created by a laser pulse with $a_0(z)$ (dashed red line) in a gas jet with gaussian density distribution (shaded green area). Bottom: Initial electron beam phase space and spectrum (blue) and chirp-compensated beam (red).}
\label{Fig4}
\end{figure}

\section{Towards percent-level energy spread}

In the future, it would be even more attractive if density tailoring could provide complete phase space control, for instance to reach sub-percent level energy spread. However, such performance is difficult to reach in the non-linear blowout regime, both in simulations and experiments. This is because fine tuning of the wakefields is difficult because of the complex interplay of plasma density and laser evolution. Furthermore, the electron blowout creates sawtooth-like, negative-gradient wakefields, which cannot compensate for positive-chirp. The latter occurs for example in shock-front injection and limits the minimal energy spread of this otherwise high-quality injection technique. Both of these drawbacks can be addressed when operating in the (quasi-)linear wakefield regime. 

Using a setup similar to a laser-plasma lens \cite{Lehe:2014efa}, the remaining part of the accelerating laser pulse or a second laser pulse could be used to create a (quasi-)linear wakefield in a subsequent, second gas target that solely serves the purpose of de-chirping the beam. In the case of a single pulse, the intensity of latter can be adjusted by tuning the distance between the accelerator and de-chirping stage, while the effective accelerating fields can be fine tuned with the density profile. As an illustration of this scheme's potential, Fig.6 shows results based on fluid model calculations. 

To model the performance of chirp compensation in the quasi-linear regime, we rely on the 1-D fluid model. The use of such a model is justified for laser intensities $a_0\ll1$, which are typically reached a few Rayleigh lengths behind the wakefield accelerator. Instead of analytically solving the wave equation in the linear regime~\cite{Lehe:2014efa}, we numerically solve the ODE
$$ \frac{\partial^2}{\partial \xi^2} \Phi =  \frac12\left( \frac{1+a^2}{(1+\Phi)^2} - 1 \right)k_p^2,$$
using a finite difference approach. Here $\Phi$ is the potential, $a$ is the laser potential and $\xi$ are the co-moving coordinates. We furthermore assume that the laser does not self-focus, but evolves as a gaussian beam. The dephasing of the electron beam is taken into account using the group velocity of a laser in a cold, underdense plasma ($v_g/c\simeq 1-n_e/2n_c$) and a highly relativistic electron bunch ($v_e/c\simeq 1$).
As for experiments on laser-plasma lenses~\cite{Thaury:2015cg}, we assume that longitudinal density tailoring is performed using a sonic gas jet. The wakefield is then calculated along the laser propagation using $n_e(z)$ and $a_0(z)$. Based on the input beam parameters, we then optimize the defocusing length, the gas jet length and the peak density to get the lowest possible energy spread. Finding the optimal parameters can take hundreds of iterations, which is the main reason for the use of a fluid model in contrast to PIC codes in this study. Beam-loading and self-focusing of the electron beam could potentially alter the results, which we will address in future studies.

For the de-chirping stage a gaussian gas density profile of variable width and peak density is assumed, which is typical for targets based on sonic gas jets. Starting from a beam with a linear chirp of $\alpha_{\mbox{\scriptsize{initial}}}=-0.6$ MeV/fs, the chirp can be almost entirely compensated ($\alpha_{\mbox{\scriptsize{final}}}=-0.03$ MeV/fs) and  the rms energy spread is lowered to 0.4\% (1.1\% FWHM). Additionally, the higher order chirp and with it the longitudinal emittance are reduced. The results indicate that such a chirp compensation in a longitudinally tailored plasma could be an alternative to other proposals, like chirp mitigation in density modulated plasmas \cite{Brinkmann:2017kh}, in order to reach sub-percent level energy spread beams in laser-wakefield accelerators.

\section{Conclusions}

In conclusion, we have presented results on energy chirp compensation and energy spread reduction in density-tailored laser wakefield accelerators. The results extend the laser-plasma lensing and rephasing concepts to the production of high-quality electron beams. Experimentally, we demonstrated an energy spread reduction of a broadband electron beam to less than 10 percent, while maintaining a high charge of about 120 pC. The method facilitates the production of highly charged bunches of monoenergetic electrons and is simple to implement in existing setups using either gas jets or double compartment gas-cells \cite{Kononenko:2016cz}. This kind of beams is of immediate interest for laser-driven X-ray sources, such as Compton sources \cite{Dopp:2016ki,Khrennikov:2015gxa}, and free-electron lasing experiments \cite{Maier:2012eq,Couprie:2013km}, which would profit from the increased spectral charge density while maintaining the beam divergence. Furthermore, density tailoring in the quasi-linear regime may lead to the production of even lower, sub-percent energy spread beams. 

\vspace{5mm}
\small{ACKNOWLEDGMENTS: This project has received funding from the European Union's Horizon 2020 Research and Innovation programme under grant agreement No 730871 (project ARIES) and from the European Research Council (ERC) under grant agreement No 339128 (project X-Five). We acknowledge also the support from the Agence Nationale pour la Recherche through the projects FEMTOMAT (ANR-13-BS04-0002) and LUCELX project (ANR-13-BS04-0011). The authors gratefully acknowledge the Gauss Centre for Supercomputing e.V. (www.gausscentre.eu) for funding this project by providing computing time on the GCS Supercomputer SuperMUC at Leibniz Supercomputing Centre (www.lrz.de) under project id pn69ri.  A.D. acknowledges support by DFG through the Cluster of Excellence Munich-Centre for Advanced Photonics (MAP EXC 158) and thanks the \textsc{Osiris} consortium (IST/UCLA) for access to the \textsc{Osiris} code. F. Massimo received support by the European Union's Horizon 2020 research and innovation programme under grant agreement EuPRAXIA No. 653782.}

\end{document}